\documentclass[11pt]{revtex4-1}

\usepackage{graphicx}

\begin{document}

\title{A unified cosmic evolution: Inflation to late time acceleration}

\author{Subenoy Chakraborty\footnote {schakraborty@math.jdvu.ac.in}}
\author{Supriya Pan\footnote {span@research.jdvu.ac.in}}
\author{Subhajit Saha\footnote {subhajit1729@gmail.com}}

\affiliation{Department of Mathematics, Jadavpur University, Kolkata 700032, West Bengal, India.}


\begin{abstract}
The present work deals with a cosmological model having particle creation mechanism
in the framework of irreversible thermodynamics. In the second order non-equilibrium
thermodynamical prescription, the particle creation rate is treated as the dissipative
effect. The non-equilibrium thermodynamical process is assumed to be isentropic, and,
as a consequence, the entropy per particle is constant, and, hence, the dissipative
pressure can be expressed linearly in terms of the particle creation rate in the
background of the homogeneous and isotropic flat FLRW model. By proper choice of
the particle creation rate as a function of the Hubble parameter, the model shows
the evolution of the universe starting from the inflationary scenario to the present
accelerating phase, considering the cosmic matter as normal perfect fluid with
barotropic equation of state.

\end{abstract}

\maketitle

\section{Introduction}
In cosmology, usually homogeneous and isotropic flat Friedmann-Lemaitre-Robertson-Walker (FLRW)
model is considered due to its agreement with inflation and the cosmic microwave background (CMB)
observations. Here, the only dissipative phenomenon is in the form of bulk viscous pressure which
may occur either due to coupling of different components of the cosmic substratum
\cite{Weinberg1, Straumann1, Schweizer1, Udey1, Zimdahl1}, or, due to non-conservation of (quantum)
particle number \cite{Zeldovich1, Murphy1, Hu1}. In the present work, second option is only
considered and for simplicity of calculations, isentropic (i.e., adiabatic) particle
production \cite{Prigogine1, Calvao1} of perfect fluid particles is considered, and,
consequently, there is a simple linear relationship between particle production rate, and,
 the viscous pressure. However, it is to be noted that still there is entropy production
due to enlargement of the phase space (due to increase in the number of fluid particles, and, also,
 due to the expansion of the universe in the present model) of the system.

Further, in the context of non-equilibrium thermodynamical prescription, second
order deviations from equilibrium (following Israel and Stewart \cite{Israel1, Israel2})
are considered in order to eliminate the drawbacks of the 1st order Eckart theory related
to causality and stability. As a result, entropy flow will be related to the cosmological
particle production of isentropic nature, and, the bulk viscous pressure will become a
dynamical degree of freedom having causal evolution equation. Furthermore, it is possible
to eliminate bulk viscous inflation without particle production (as used by many authors),
and, thus, one may have a model universe starting from a de Sitter phase which gradually
evolves to standard FLRW model. The present work will be an attempt to incorporate the
present accelerating phase within this causal second order theory.

Usually, to explain the recent accelerated expansion of the Universe as predicted by
Supernovae Ia and complementary observations \cite{Riess1, Perlmutter1}, there are two
common approaches $-$ one within the framework of Einstein gravity by introducing an
unknown type of matter component with a large negative pressure (dark energy), and,
secondly, by modifying Einstein gravity theory, and interpreting the extra geometric
terms as hypothetical matter component to explain the present accelerating phase.
However, both the approaches are concentrated only to explain the recent observational
predictions $-$ there is no concern about the past or future evolution of the universe.
In this context, the present work is an attempt not only to explain the recent
observations, but also the past evolution of the universe without following any
one of the above mentioned conventional approaches.\\

The paper is organized as follows: Section \ref{PCR} deals with a brief review of the
non-equilibrium thermodynamics in cosmology, section \ref{Phenomenological} describes
the explicit solutions of the cosmological parameters corresponding to a unique particle creation rate.
In section \ref{Cosmographic}, we have analyzed the cosmographic parameters associated with 
our model. Section \ref{Field}, we have described the particle productions in the language of
field theory. In section \ref{Hawking}, we present an associated Hawking like radiation. Finally,
we finish our discussions in section \ref{Summary}.

\section{Particle Creation Mechanism and Non-equilibrium Thermodynamics: Basic Equations}
\label{PCR}
In a closed thermodynamical system, suppose there are $N$ particles having internal
energy $E$. Then the first law of thermodynamics which is essentially the conservation
of internal energy reads

\begin{equation}
dE=dQ-pdV,
\end{equation}

where $dQ$ is the amount of heat received by the system in time $dt$, and as
usual $p$ is the thermodynamic pressure, $V$ is any co-moving volume.
The above energy conservation relation can be rewritten as Gibb's equation

\begin{equation} \label{ge}
Tds=d\tilde{q}=d\left(\frac{\rho}{n}\right)+pd\left(\frac{1}{n}\right),
\end{equation}

where `$s$' is the entropy per particle (specific entropy), $T$ is the temperature of
the fluid, $n=N/V$ is the particle number density, and $d\tilde{q}= dQ/N$ is the heat
per unit particle. It should be noted that the above Gibb's equation also holds for
open thermodynamical system, i.e., when the particle number is not conserved
\cite{Prigogine1, Calvao1, Lima0}.

As mentioned earlier, we consider spatially flat FLRW model of the universe as an
open thermodynamical system, and, non-equilibrium nature comes into picture due to the
particle creation mechanism. Thus, the Einstein's field equations take the form

\begin{equation} \label{efe}
3H^2=\kappa \rho,~~~~~~\mbox{and,}~~~~~~2\dot{H}=-\kappa(\rho +p+\Pi),
\end{equation}

where $\kappa =8\pi G$ is the Einstein's gravitational constant. For a relativistic
fluid with dissipation in the form of bulk viscosity has the energy-momentum tensor

\begin{equation}
T_{\mu \nu}=(\rho +p+\Pi)u_{\mu}u_{\nu}+(p+\Pi)g_{\mu \nu},
\end{equation}

and the conservation equation ${T^{\mu \nu}}_{;\nu} =0$ (Bianchi's identity) reads

\begin{equation} \label{ece}
\dot{\rho}+\theta(\rho +p+\Pi)=0,
\end{equation}

where the bulk viscous pressure $\Pi$ is related to the entropy production.

Further, in an open thermodynamical system, the non-conservation of
fluid particles is reflected by the equation

\begin{equation} \label{pnce}
{N^\mu}_{;\mu} \equiv \dot{n}+\theta n= n \Gamma~,
\end{equation}

where $\Gamma$ denotes the rate of change of the particle number ($N=na^3$)
in a co-moving volume $a^3$. $N^\mu =nu^\mu$ is the particle flow vector, $n$
is the particle number density, $u^\mu$ is the unit time-like vector ($4-$velocity),
$\theta ={u^{\mu}}_{;\mu}$ is the expansion of the congruence of time-like geodesics
and notationally $\dot{n}=n_a u^a$. So, creation and annihilation of particles is
characterized by the sign of $\Gamma$ (creation: $\Gamma >0$ and annihilation: $\Gamma <0$).
Further, a non-zero $\Gamma$ is dynamically equivalent to an effective bulk pressure
\cite{Hu1, Maartens1, Barrow1, Barrow2, Barrow3, Barrow4, Peacock1} of the fluid, and, hence,
non-equilibrium thermodynamics comes into consideration. However, Lima et al. \cite{Lima1}
showed that such scalar processes (bulk viscosity and matter creation) are not equivalent
from a thermodynamic viewpoint $-$ only the dynamic behavior can simply be demonstrated
in the case of ``adiabatic'' particle creation as follows:

Using the above conservation equations (\ref{ece}) and (\ref{pnce}) into the Gibb's
relation (i.e., Eq. (\ref{ge})), one obtains the entropy variation per particle as

\begin{equation}
\dot{s}=-\frac{\theta}{nT}\left[\Pi +\frac{\Gamma}{\theta}(\rho +p)\right].
\end{equation}

For simplicity, if the thermal process is assumed to be isentropic (i.e., adiabatic)
then the entropy per particle remains constant (in contrast to dissipative process),
i.e., $\dot{s}=0$ and we have

\begin{equation} \label{ic}
\Pi =-\frac{\Gamma}{\theta}(\rho +p).
\end{equation}

The above relation shows that a non-vanishing $\Gamma$ will produce an effective
bulk pressure on the thermodynamic fluid and non-equilibrium thermodynamics comes
into the picture. In other words, a dissipative fluid is equivalent to a perfect
fluid having a non-conserved particle number. It should be noted that there is
entropy production only due to the enlargement of the phase space of the system.
Further, one may note that, this effective bulk pressure does not correspond to
conventional non-equilibrium phase, rather a state having equilibrium properties
as well (but not the equilibrium era with $\Gamma =0$).

In the framework of second order non-equilibrium thermodynamics due to Israel
and Stewart \cite{Israel1}, the entropy flow vector ($S^a$) has the expression

\begin{equation} \label{efv}
S^{a}= sN^{a}-\frac{\tau {\Pi}^2}{2\zeta T}u^{a},
\end{equation}

where $\zeta$ is the coefficient of bulk viscosity and $\tau$ represents the time
of relaxation. Now, using the conservation equations (\ref{ece}) and (\ref{pnce}),
one obtains the entropy production density from the above Eq. (\ref{efv}) as

\begin{equation}
T{S^{a}}_{;a}=-n\mu \Gamma -\Pi \left[\theta +\frac{\tau \dot{\Pi}}{\zeta}+\frac{1}{2}\Pi T{\left(\frac{\tau}{\zeta T}u^{a}\right)}_{;a}\right],
\end{equation}

where $\mu =\left(\frac{\rho +p}{n}\right)- Ts$ is the chemical potential.
Now, for the validity of the second law of thermodynamics
(i.e., ${S^a}_{;a} \geq 0$) one can choose the ansatz for bulk viscous pressure as

\begin{equation}
\Pi =-\zeta \left[\theta +\frac{\tau}{\zeta}\dot{\Pi}+\frac{1}{2}\Pi T\left(\frac{\tau}{\zeta T}u^{a}\right)_{;a}+\frac{\mu n \Gamma}{\zeta \Pi}\right].
\end{equation}

As a result, $\Pi$ becomes a dynamical variable with an inhomogeneous evolution equation

\begin{equation}
{\Pi}^2+\tau \Pi \dot{\Pi}+\frac{1}{2}\zeta {\Pi}^2 T{\left(\frac{\tau}{\zeta T}u^{a}\right)}_{;a}+\zeta \Pi \theta =-\zeta \mu n\Gamma .
\end{equation}

Here, the chemical potential $\mu$ may act as an effective symmetry-breaking
parameter in relativistic field theories. Also, the evolution equation becomes
linear first order in nature in absence of chemical potential.

In this context, it is relevant to mention that the basic physical difference
between the noncausal and the causal theory is the introduction of the time of
relaxation (in the later one). As a result, in causal theory $\Pi$ decays to zero
after $\Gamma$ has been switched off (assuming non-vanishing $\Gamma$ produces the
effective viscous pressure). Moreover, if the above second order theory is isentropic
in nature then the entropy production density simplifies to

\begin{equation}
{S^a}_{;a}=-\Pi \left[\frac{\Pi}{2}\left(\frac{\tau}{\zeta T}u^{a}\right)_{;a}+\frac{\tau}{\zeta T}\dot{\Pi}-\frac{ns\Gamma}{\Pi}\right].
\end{equation}

Further, due to the isentropic condition (\ref{ic}), the evolution of the
relevant thermodynamical variables are \cite{Zimdahl2}

\begin{equation} \label{eeqs}
\dot{\rho}=-(\theta -\Gamma)(\rho +p)~,~~\dot{p}=-c_s^{2}(\theta -\Gamma)(\rho +p)~,~~\frac{\dot{n}}{n}=-(\theta -\Gamma)~,~~\frac{\dot{T}}{T}=-(\theta -\Gamma)\frac{\partial p}{\partial \rho},
\end{equation}

where $c_{s}^{2}=\left(\frac{\partial p}{\partial \rho}\right)_{ad.}$ is the
square of the adiabatic sound velocity \cite{Zimdahl2}.

\section{Phenomenological Choice of Particle Creation Rate and Cosmic Evolution}
\label{Phenomenological}
To describe the cosmic evolution (for a given particle creation rate as a
function of the Hubble parameter), one can eliminate the effective bulk
pressure $\Pi$ from the Einstein field equations (\ref{efe}) using the
isentropic condition (\ref{ic}) to obtain

\begin{equation} \label{g3h}
\frac{\Gamma}{3H}=1+\frac{2}{3\gamma}\left(\frac{\dot{H}}{H^2}\right).
\end{equation}

On the other hand, this equation can also be considered as the determining
equation for the particle creation rate $\Gamma$ from the given cosmic evolution.
Also, the deceleration parameter $q$ takes the form

\begin{equation}
q=-1+\frac{3\gamma}{2}\left(1-\frac{\Gamma}{3H}\right).
\end{equation}

So, in the present context, the cosmic history is characterized by the
fundamental physical quantities, namely, the expansion rate $H$ and the
energy density $\rho$, and, as a result, the gravitational creation rate
$\Gamma$ can be defined in a natural way.

In earlier studies \cite{Lima1, Zimdahl2, Abramo1, Gunzig1998, Lima2, Basilakos1},
the particle creation rate $\Gamma$ has been chosen for different phases of the
evolution from thermodynamical viewpoint. As $\Gamma$ should be greater than $H$
in the very early universe so that the created radiation may be considered as a
thermalized heat bath. Hence, at the very early Universe, $\Gamma$ is chosen to
be proportional to $H^2$ \cite{Abramo1, Gunzig1998} (i.e., $\Gamma \propto \rho$)
and the corresponding cosmological solution \cite{Lima1, Lima2, Basilakos1, Zimdahl2}
shows a smooth transition from the inflationary scenario to the radiation era. Also
for this adiabatic production of relativistic particles, the energy density scales
as $\rho _r \propto T^4$, i.e., black body radiation \cite{Lima2, Basilakos1}.

Furthermore, it has been recently shown \cite{Saha1, Pan1} that, $\Gamma \propto H$
and $\Gamma \propto 1/H$ describe respectively the intermediate matter dominated era
(starting from radiation) and the transition from matter dominated era to late time
acceleration. The scale factor, the Hubble parameter and the thermodynamical parameters
are shown to be continuous across the transition points (i.e., the epochs from the
inflationary era to the radiation era and from the matter dominated era to the late
time acceleration). Also, $\Gamma =\Gamma _0$ has been shown \cite{Chakraborty1} to
correspond the emergent scenario. Subsequently, in another work \cite{Chakraborty2},
a linear combinations of all these choices, i.e., $\Gamma =\Gamma _0+ lH^2+ mH+ n/H$
has been examined to describe the whole evolution of the Universe. Although no exact
analytic solution is possible for this choice of $\Gamma$ (from Eq. (\ref{g3h})),
still the graphical representation of the deceleration parameter has shown the whole
evolution of the universe starting from an early inflationary epoch to the present
accelerating phase, and, the model predicts a possible transition from present
accelerating stage to decelerating phase again in the future.

\begin{figure}
\begin{minipage}{0.45\textwidth}
\includegraphics[width= 0.85\linewidth]{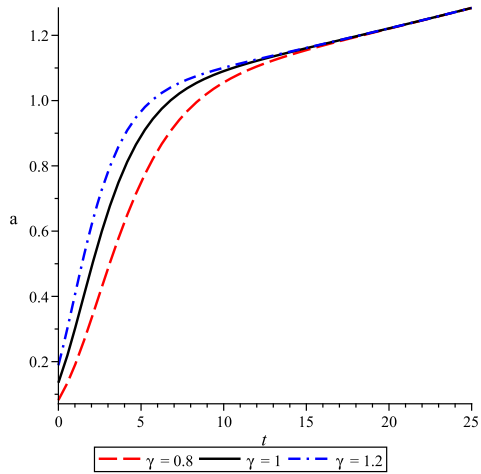}\\
\end{minipage}
\caption{This figure describes the scale factor over the cosmic time.}
\end{figure}

\begin{figure}
\begin{minipage}{0.45\textwidth}
\includegraphics[width= 0.85\linewidth]{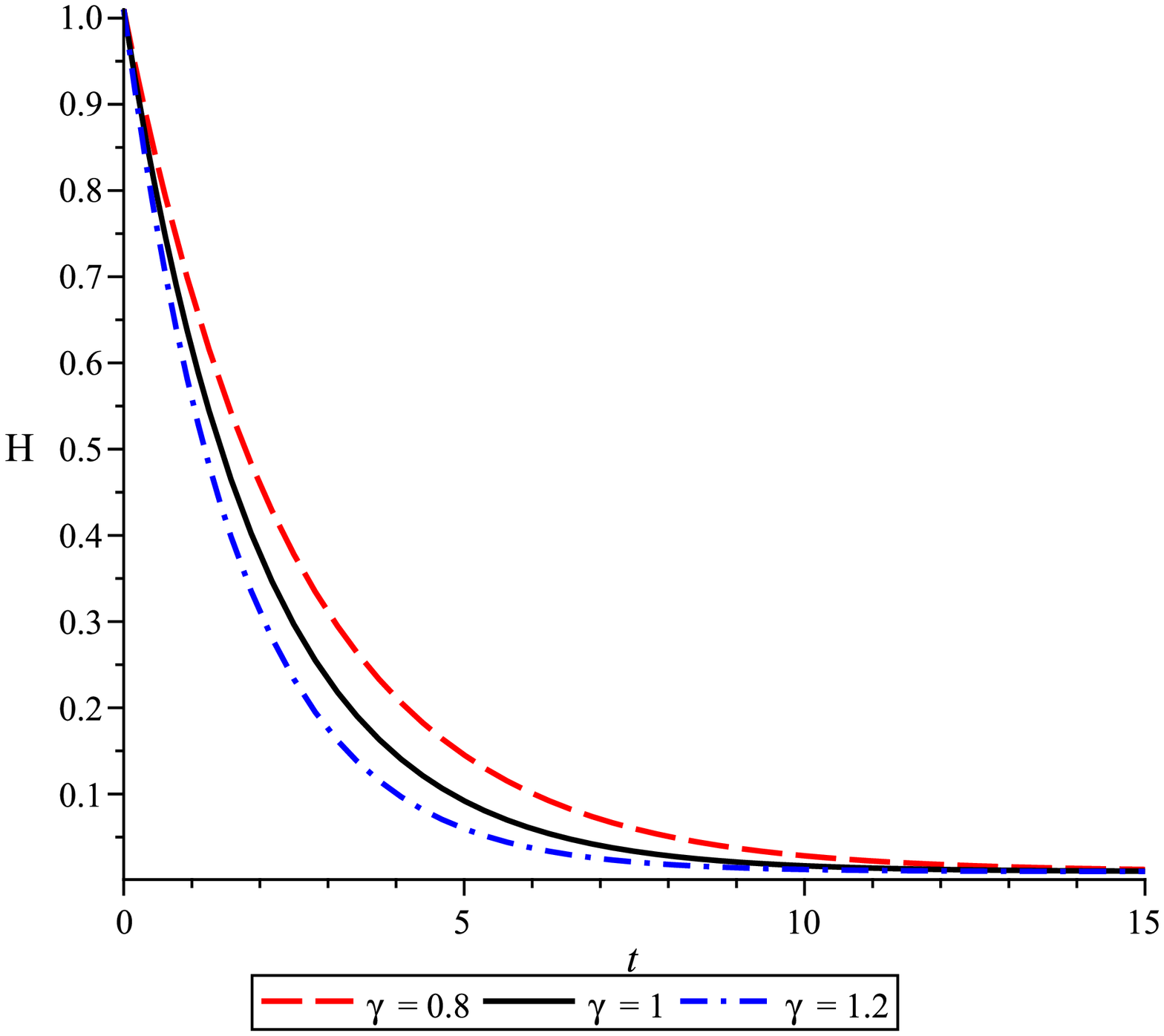}
\caption{The figure shows the behavior of the\\ Hubble parameter over the cosmic time.}
\end{minipage}
\begin{minipage}{0.45\textwidth}
\includegraphics[width= 0.85\linewidth]{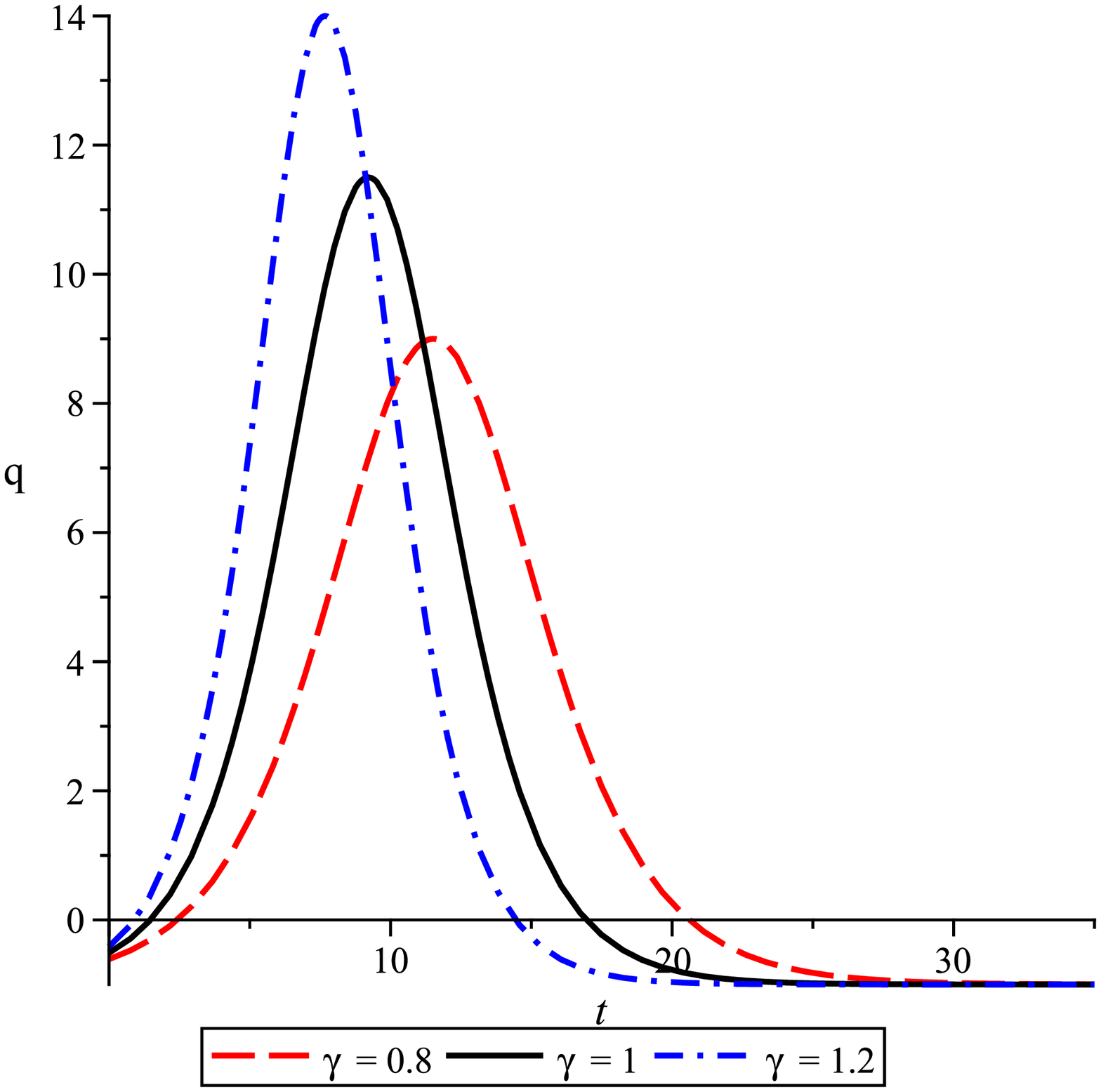}
\caption{This displays the entire cosmic history from inflation to present accelerating phase of the universe.}
\end{minipage}
\end{figure}

With this background in mind, the present work is a partial modification of
the above general choice with an aim to have an exact analytic solution so that
more sophisticated cosmic study, namely, the cosmographic analysis can be done.
By choosing the coefficients appropriately (which will be clear subsequently)
the form of $\Gamma$ is taken as

\begin{equation} \label{gamma}
\Gamma =-\mu ^2+3H+\frac{\alpha ^2}{H}
\end{equation}

with $\mu$ and $\alpha$ as real constants. Now substituting this $\Gamma$ into the evolution equation (\ref{g3h}), the explicit solution reads as

\begin{eqnarray} \label{soln}
a &=& a_0 e^{\left(\frac{\alpha ^2}{\mu ^2}t\right)} exp\left[-\frac{2}{\gamma \mu ^4}exp\left\lbrace -\frac{\mu ^2 \gamma}{2}(t-t_0)\right\rbrace \right],\\
H &=& \frac{\alpha ^2}{\mu ^2}+\frac{1}{\mu ^2} exp\left[-\frac{\mu ^2 \gamma}{2}(t-t_0)\right],
\end{eqnarray}

where $a_0$ and $t_0$ are the constants of integration. The graphical representation
of the cosmic evolution, namely, the scale factor $a$, the Hubble parameter $H$, and
the deceleration parameter $q$ are presented in FIGs. (1)--(3) respectively for various
choices of $\gamma$, the equation of state parameter for the cosmic fluid. The diagramatic
representation of $q$ shows two transitions of $q$ (from accelerating phase to deceleration,
and, then again acceleration) which indicates that the present model of the Universe describes
the evolution from the inflationary scenario to the present late time acceleration through the
decelerated matter dominated era. Thus, we have a complete cosmic history after the big bang
till today. It is interesting to note that as the cosmic time becomes very large, we have
the $\Lambda$CDM model:

\begin{eqnarray}
a &\simeq & a_0 e^{H_0 t},~H \simeq H_0=\alpha ^2/\mu ^2,~q \simeq -1,~,~\rho \simeq 3H_{0}^{2}=\Lambda =-p,\label{LambdaCDM1}
\end{eqnarray}

and it agrees with the recent Planck data set \cite{Planck2015}.
Now, corresponding to the cosmological solution (i.e., Eqns. (\ref{gamma})
and (\ref{soln})) of the present model, the relevant thermodynamical
parameters evolve as (see Eq. (\ref{eeqs}))

\begin{eqnarray}
T &=& T_0 \left[\frac{\alpha ^2}{\mu ^2}+\frac{1}{\mu ^2}exp\left\lbrace -\frac{\mu ^2 \gamma}{2}(t-t_0)\right\rbrace \right]^{\left(\frac{\gamma -1}{\gamma}\right)} \nonumber \\
n &=& n_0 \left[\frac{\alpha ^2}{\mu ^2}+\frac{1}{\mu ^2}exp\left\lbrace -\frac{\mu ^2 \gamma}{2}(t-t_0)\right\rbrace \right]^{\left(\frac{1}{\gamma}\right)} \nonumber \\
S &=& S_0 e^{\left(\frac{\alpha ^2}{\mu ^2}t\right)}exp\left[-\frac{2}{\gamma \mu ^4}exp\left\lbrace -\frac{\mu ^2 \gamma}{2}(t-t_0)\right\rbrace \right]\left[\frac{\alpha ^2}{\mu ^2}+\frac{1}{\mu ^2}exp\left\lbrace -\frac{\mu ^2 \gamma}{2}(t-t_0)\right\rbrace \right]^{\left(\frac{1}{\gamma}\right)},
\end{eqnarray}

where $S_0$ is some constant, and, $T_0$, $n_0$ are integration constants.
We have plotted these thermodynamical parameters in FIGs. (4)-(6) for
various choices of $\gamma$. It should be mentioned that if $\gamma =2$,
i.e., $p=\rho$ (relativistic fluid), then $T \propto \rho ^{\frac{1}{4}}$,
which represents the usual black body radiation. Note that, in the $\Lambda$CDM
limit the above thermodynamical parameters become

\begin{equation}
T \longrightarrow T_0 \left(\frac{\alpha ^2}{\mu ^2}\right)^{\left(\frac{\gamma -1}{\gamma}\right)}~,~~~~~~n \longrightarrow n_0 \left(\frac{\alpha ^2}{\mu ^2}\right)^{\left(\frac{1}{\gamma}\right)}~,~~~~~~S \longrightarrow S_0 e^{H_0 t} \left(\frac{\alpha ^2}{\mu ^2}\right)^{\left(\frac{1}{\gamma}\right)}, \label{LambdaCDM2}
\end{equation}

which shows that although the temperature and the number density become constant,
but the entropy in a co-moving volume evolves as the scale factor. Lastly, we note
that, in the above limit, the particle creation rate becomes constant:
$\Gamma \longrightarrow \Gamma _0 =3H_0$.

\begin{figure}
\begin{minipage}{0.45\textwidth}
\includegraphics[width= 0.85\linewidth]{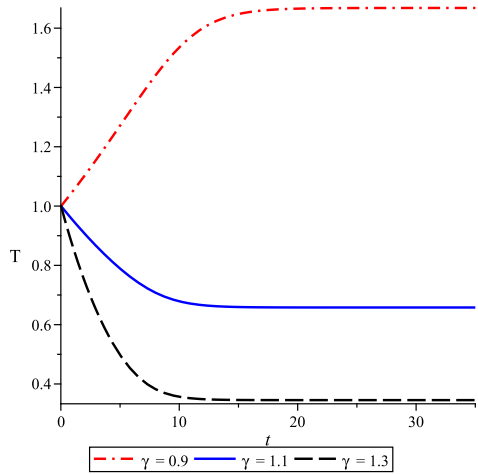}
\caption{This is the behavior of the temperature\\ over the cosmic time for three different choices\\ of $\gamma$.}
\end{minipage}
\end{figure}

\begin{figure}
\begin{minipage}{0.45\textwidth}
\includegraphics[width= 0.85\linewidth]{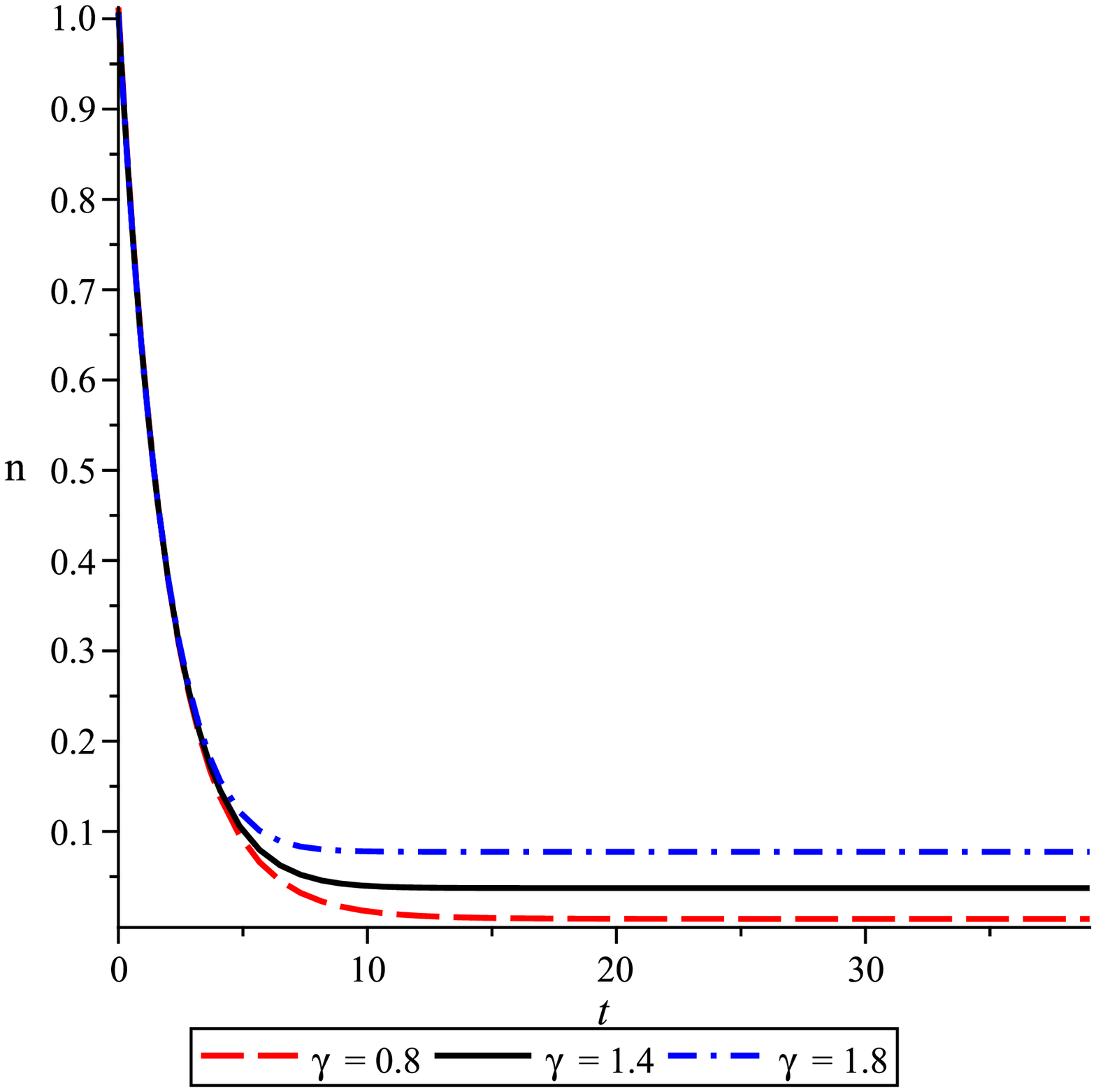}
\caption{The figure shows the variation of the\\ number density with the evolution of the\\ universe.}
\end{minipage}
\begin{minipage}{0.45\textwidth}
\includegraphics[width= 0.85\linewidth]{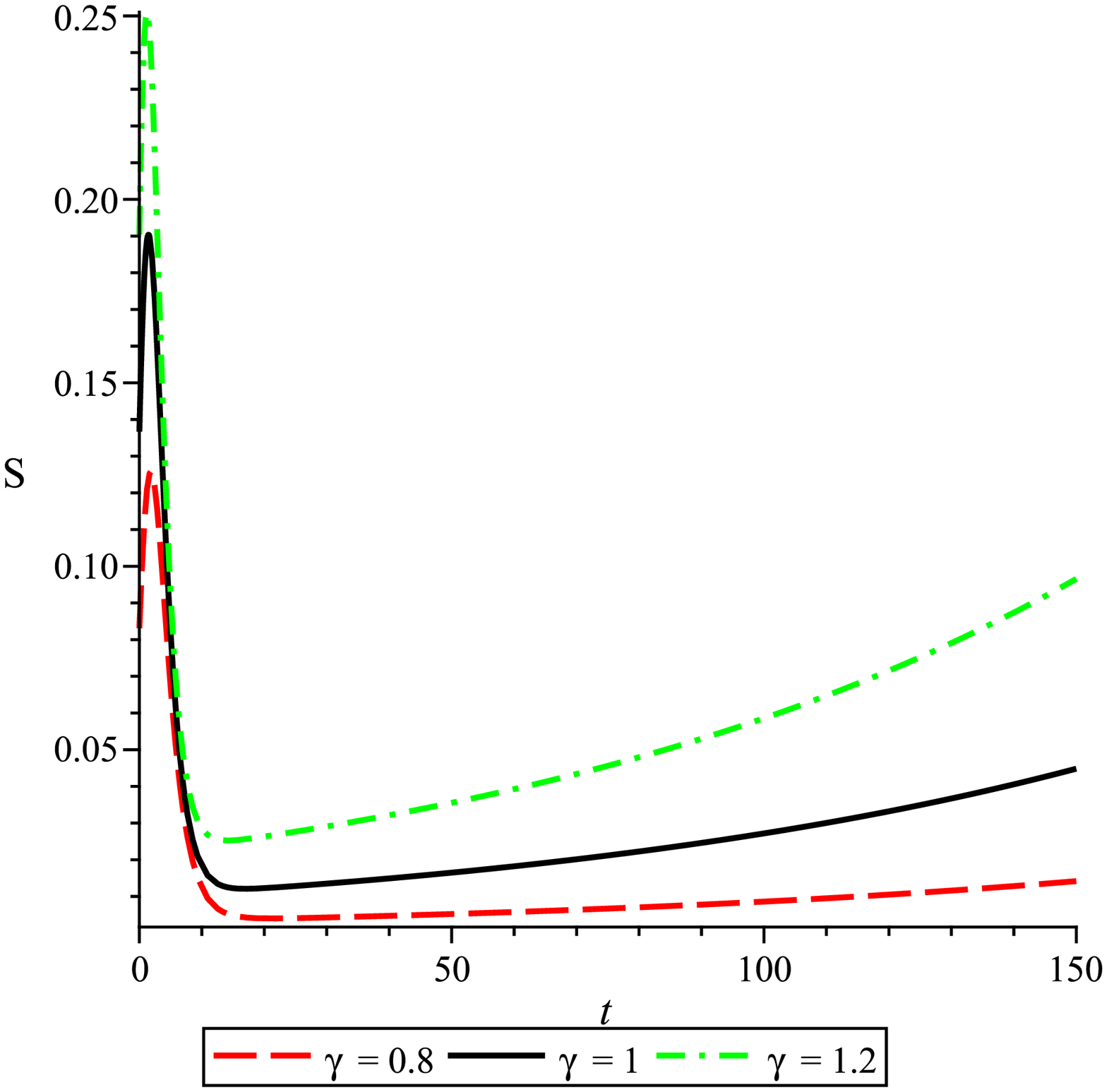}
\caption{This is the entropy variation of the\\ universe with its evolution.}
\end{minipage}
\end{figure}

\section{Cosmographic analysis and observational data}
\label{Cosmographic}
In this section, we make a comparative study of the present model of the universe
with the presently available observed data set. At first, we discuss a geometric view of
the DE models. Sahni et al. \cite{Sahni2003} first proposed this idea with two
dimensionless and model independent geometric parameters $\{r, s\}$ defined as

\begin{eqnarray}
r&=& \frac{1}{aH^3} \frac{d^3a}{dt^3},~~~\mbox{and,}~~s=\frac{r-1}{3 \left(q- \frac{1}{2}\right)}.\label{cosmographic-new1}
\end{eqnarray}

These two parameters in Eq. (\ref{cosmographic-new1}) are used to filter the observationally
supported DE models from other phenomenological DE models existing in the literature. Subsequently,
these geometric investigation was further extended by considering the Taylor series expansion of
the scale factor about the present time as

\begin{eqnarray}
\frac{a (t)}{a(t_p)}&=& 1+ H_p (t-t_p)+ \frac{1}{2!} q_p H_p ^2 (t-t_p)^2+ \frac{1}{3!} j_p H_p ^3 (t-t_p)^3+ \frac{1}{4!} s_p H_p ^4 (t-t_p)^4+O[(t-t_p)^5],\label{sp-power-series}
\end{eqnarray}

where the model independent parameters $j, s, l, m$  are known as cosmographic
parameters \cite{Visser2004, Visser2005}, and, are defined as

\begin{eqnarray}
j&=& \frac{1}{aH^3} \frac{d^3 a}{dt^3},~~s= \frac{1}{aH^4} \frac{d^4 a}{dt^4},~~l= \frac{1}{aH^5}\frac{d^5 a}{dt^5},~~\mbox{and},~~m= \frac{1}{aH^6}\frac{d^6 a}{dt^6}.\label{sp-CP-parameters}
\end{eqnarray}

Here, the suffix `$p$' stands for the value of the corresponding variable at the
present epoch ($t_p$). It should be noted that, the cosmographic parameters are individually named
as jerk ($j$) (this `$j$' is same as `$r$'
defined by Sahni et al. \cite{Sahni2003}), snap ($s$) (this `$s$' is different from
one defined by Sahni et al. \cite{Sahni2003}), lerk, and $m$
parameter \cite{Visser2004, Visser2005}.

The above cosmographic parameters (CP) can be expressed in
terms of the deceleration parameter ($q$), and, its higher derivatives
in the following way:

\begin{eqnarray}
j&=& - \frac{1}{H} \frac{dq}{dt}+ q (1+ 2 q),\label{form-j}\\
s&=& \frac{1}{H} \frac{dj}{dt}+ j- 3 (1+ q)j,\label{form-s}\\
l&=& \frac{1}{H} \frac{ds}{dt}+ s- 4 (1+q)s,\label{form-l}\\
m&=& \frac{1}{H} \frac{dl}{dt}+ l- 5(1+q)l\label{form-m}
\end{eqnarray}

\begin{figure}[h]
\begin{minipage}{0.45\textwidth}
\includegraphics[width= 1.2\linewidth]{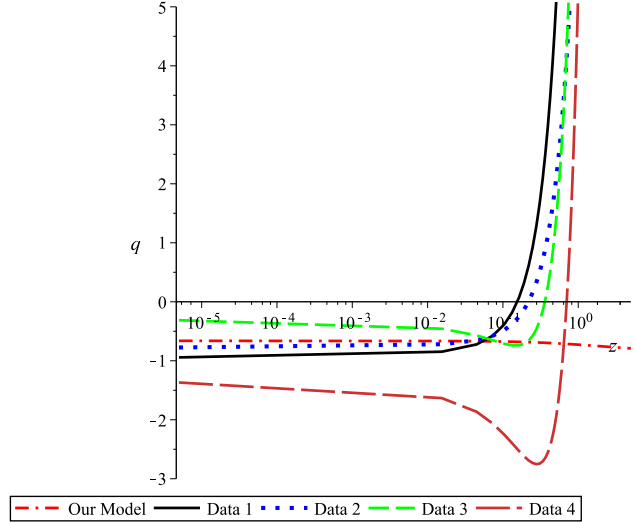}
\caption{The figure shows a comparative study of the deceleration parameter of our model with the 4 different latest observational data sets.}
\end{minipage}
\end{figure}

\begin{figure}
\begin{minipage}{0.45\textwidth}
\includegraphics[width= 1.0\linewidth]{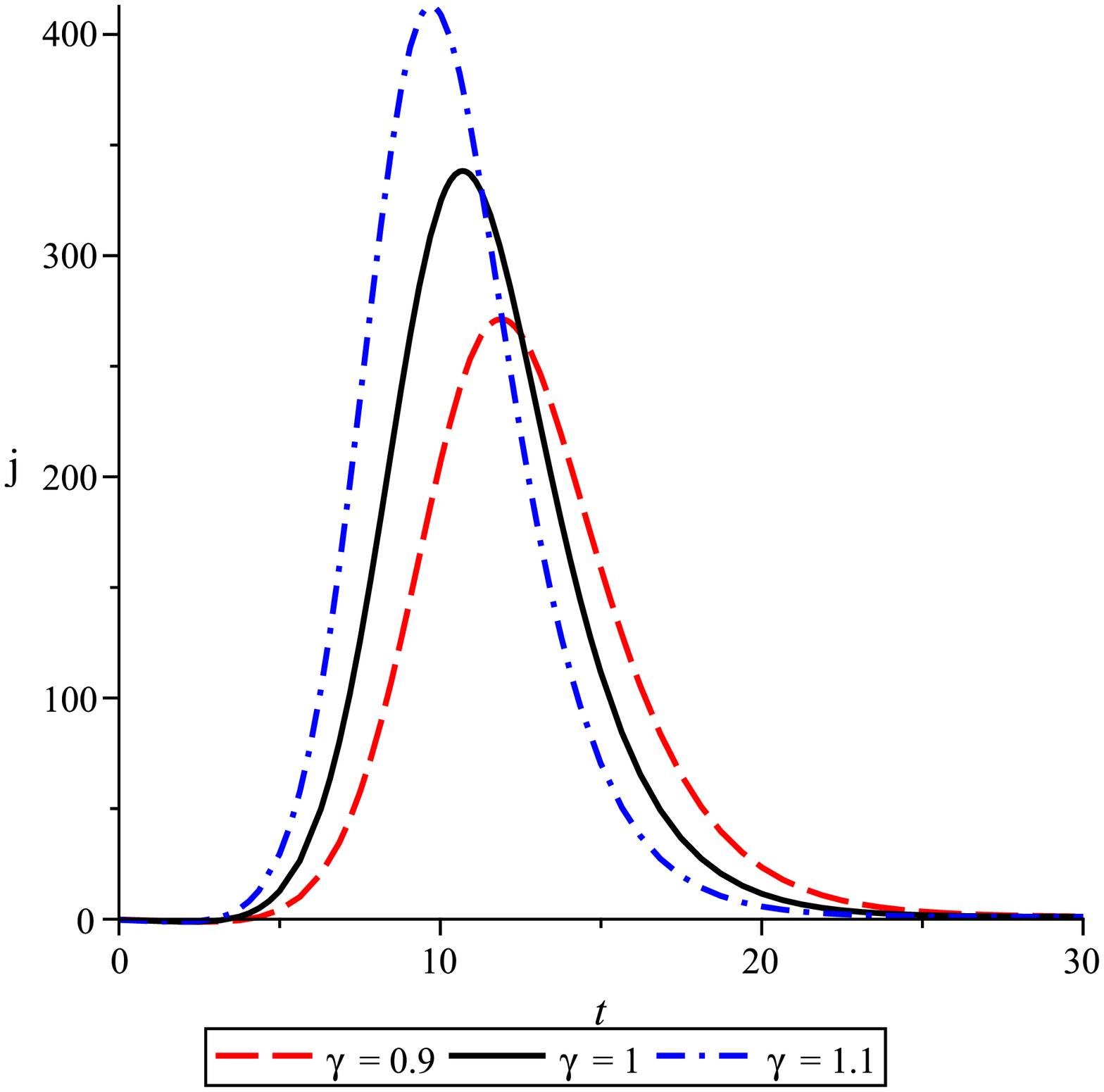}
\caption{The figures show the variation of\\ the ``jerk'' ($j$) throughout the evolution of\\ the universe.}
\end{minipage}
\begin{minipage}{0.45\textwidth}
\includegraphics[width= 0.9\linewidth]{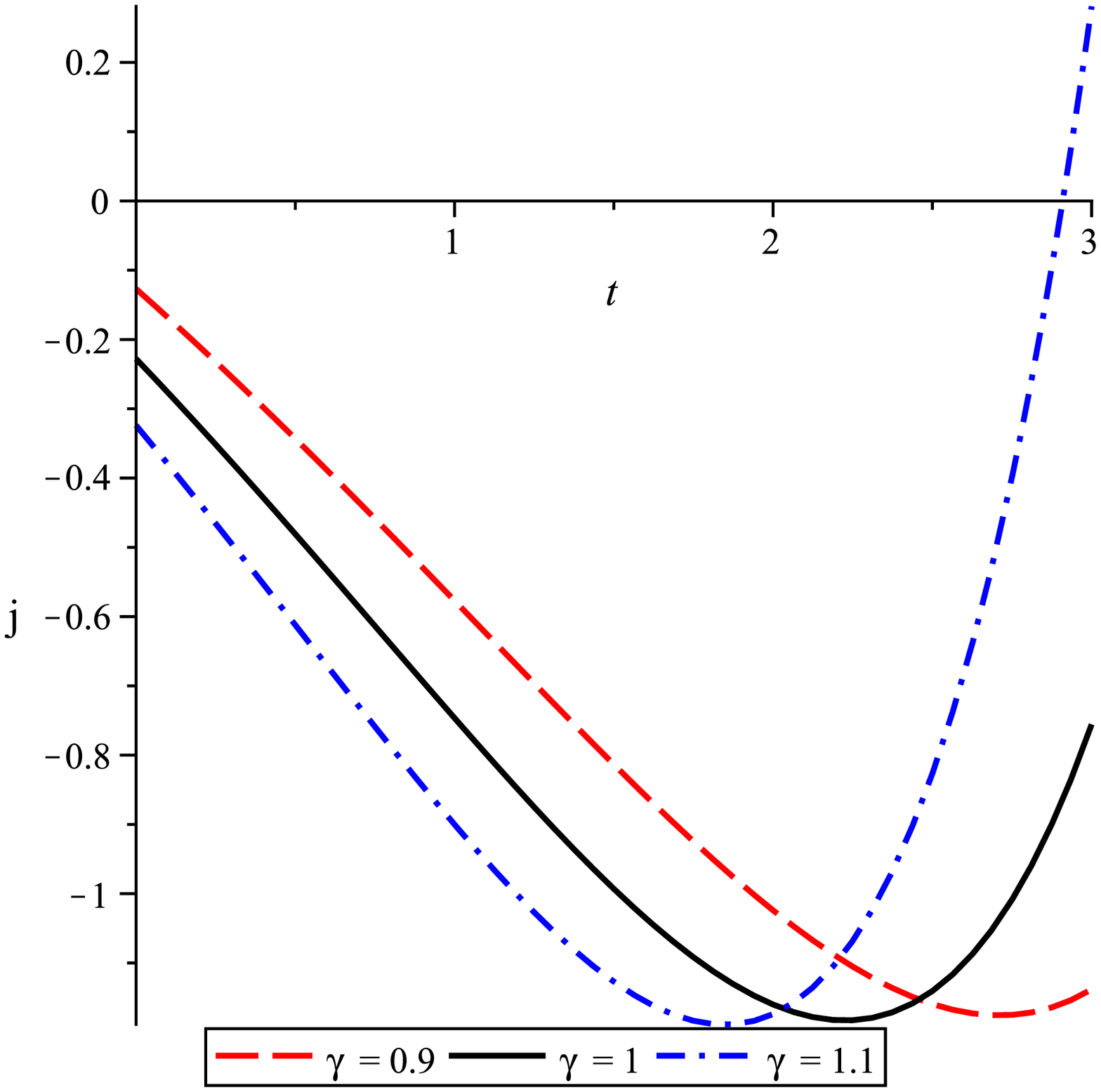}
\caption{This is just to show that during the\\ early phase of the universe, `$j$' started from\\ $-$ ve values.}
\end{minipage}
\end{figure}

Further, the deceleration parameter can be expressed in terms of the
redshift parameter $z$ ($=1/a- 1$) as

\begin{eqnarray}
q (z)&=& q_p+ (-q_p-2 q_p^2+ j_p)z+\frac{1}{2} (2 q_p+ 8 q_p^2+ 8 q_p^3- 7 q_p j_p- 4 j_p- s_p) z^2+ O(z^3).\label{sp-dp-series}
\end{eqnarray}

\begin{figure}
\begin{minipage}{0.45\textwidth}
\includegraphics[width= 1.0\linewidth]{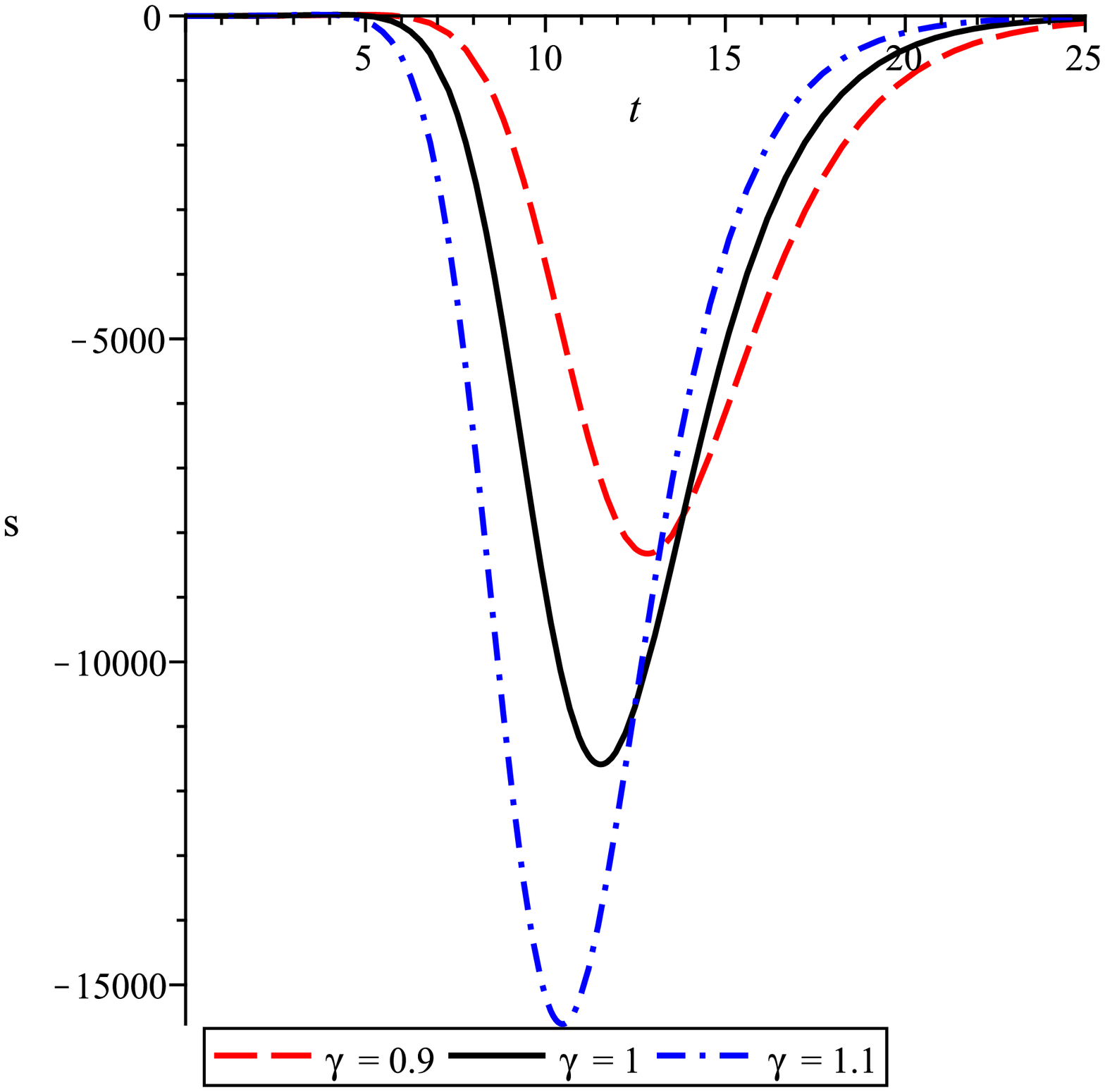}
\caption{The figures show the variation of the \\``snap'' ($s$) parameter throughout the entire\\ evolution of the universe.}
\end{minipage}
\begin{minipage}{0.45\textwidth}
\includegraphics[width= 0.9\linewidth]{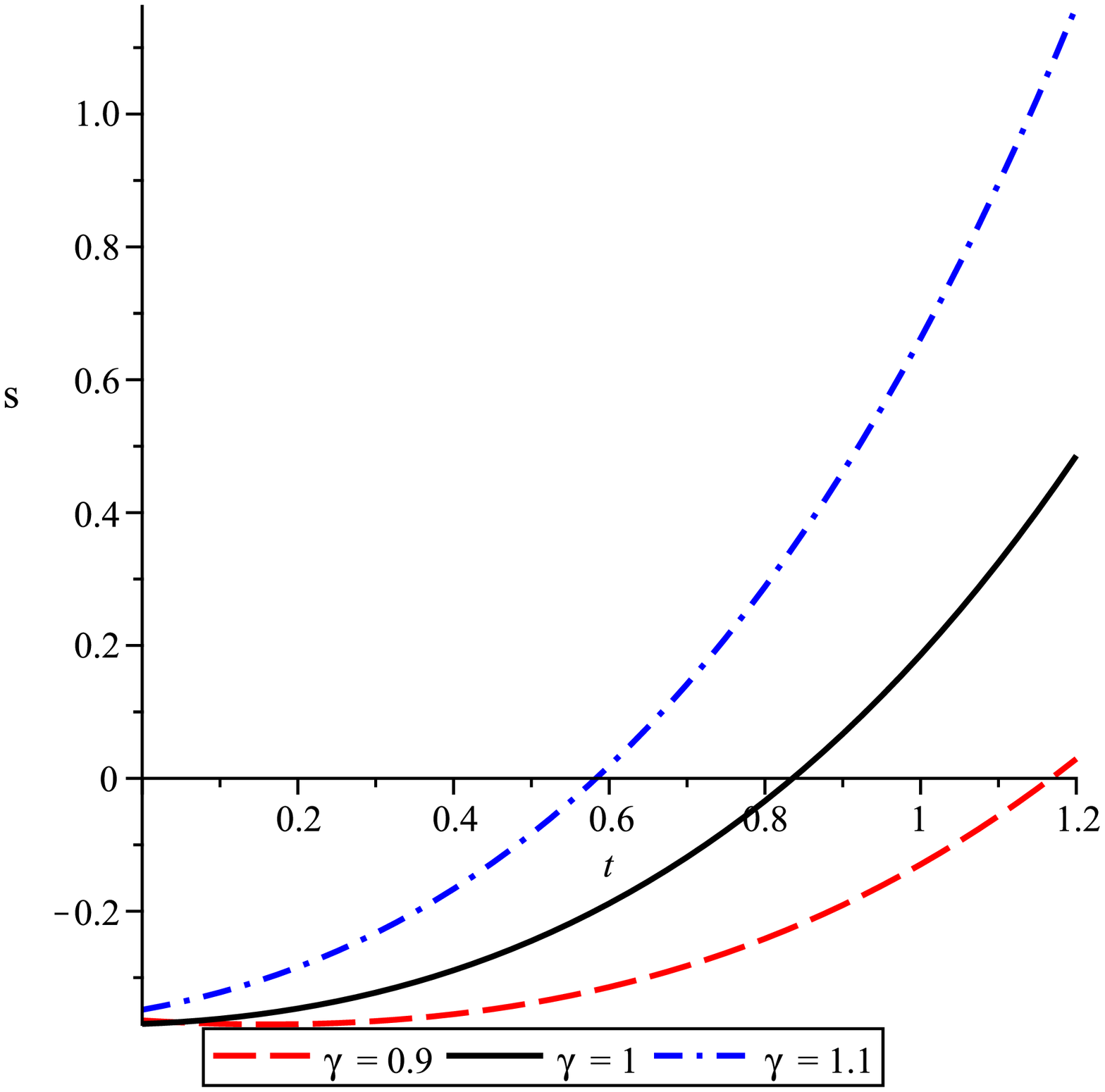}
\caption{The figures are just to show that\\ during the early phase of the universe,\\ `$s$' started from $-$ ve values.}
\end{minipage}
\begin{minipage}{0.45\textwidth}
\includegraphics[width= 0.9\linewidth]{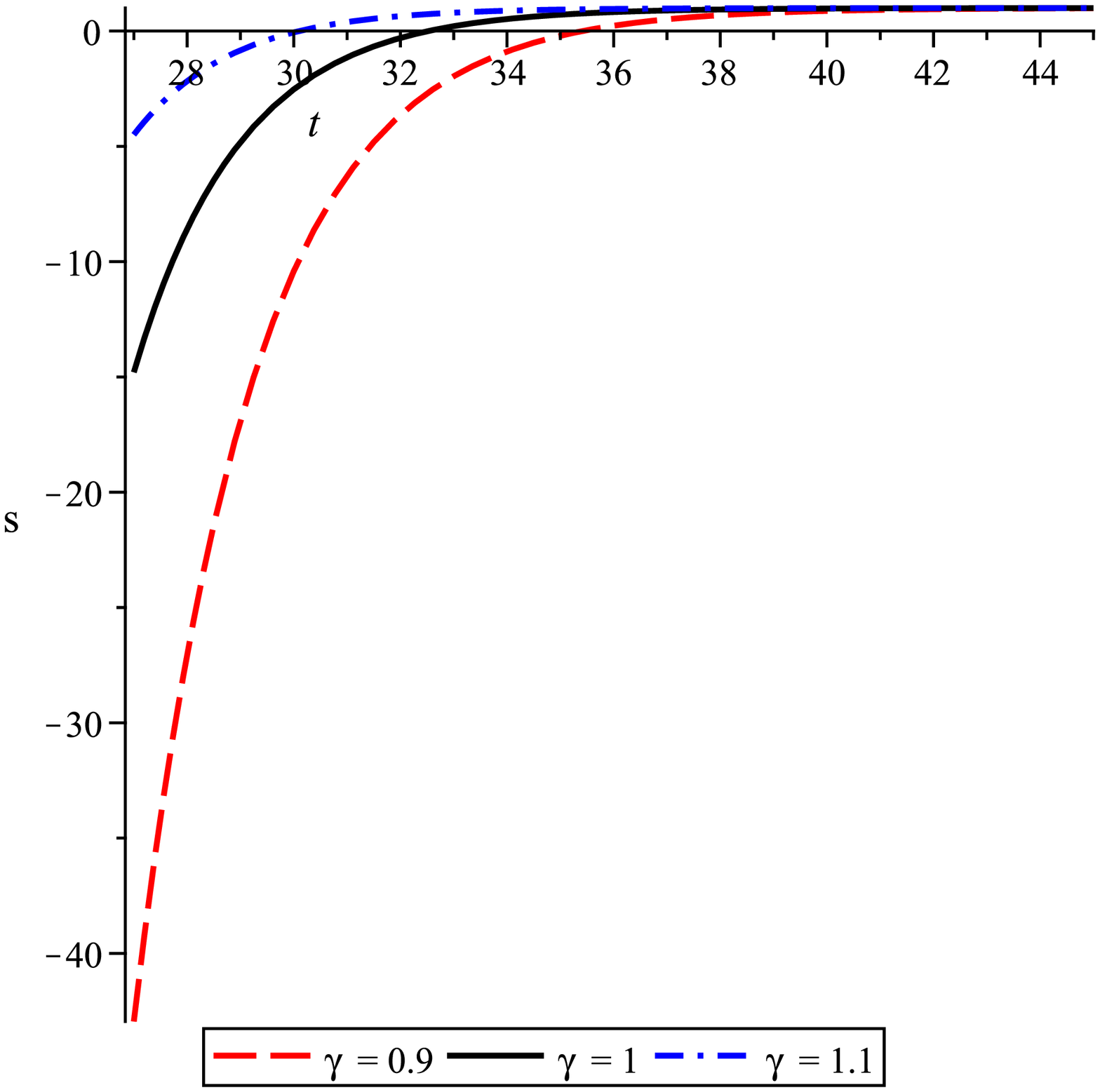}
\caption{This is just to show that when $t \longrightarrow \infty$,\\ `$s$' becomes $+$ ve.}
\end{minipage}
\end{figure}

For the observed data sets, we choose the following\\
(i) 192 Sne Ia and 69 GRBs with CPL parametrization (data 1) \cite{WDQ2009}\\
(ii) 192 Sne Ia and 69 GRBs with linear parametrization (data 2) \cite{WDQ2009}\\
(iii) Supernovae Union 2+ BAO+ OHD+ GRBs data (data 3) \cite{XW2011}\\
(iv) Supernovae Union 2+ BAO+ GRBs data (data 4) \cite{XW2011}\\

In FIG. 7, we have made a comparative study of the deceleration parameter for
the present model with those for the above four data sets. From the graph we see
that the behavior of the deceleration parameter in our model almost matches with
the 4 latest observed data sets.\\

Also, we have shown the graphical representations of the above four CP parameters
for different choices of $\gamma$.

\begin{figure}
\begin{minipage}{0.45\textwidth}
\includegraphics[width= 1.0\linewidth]{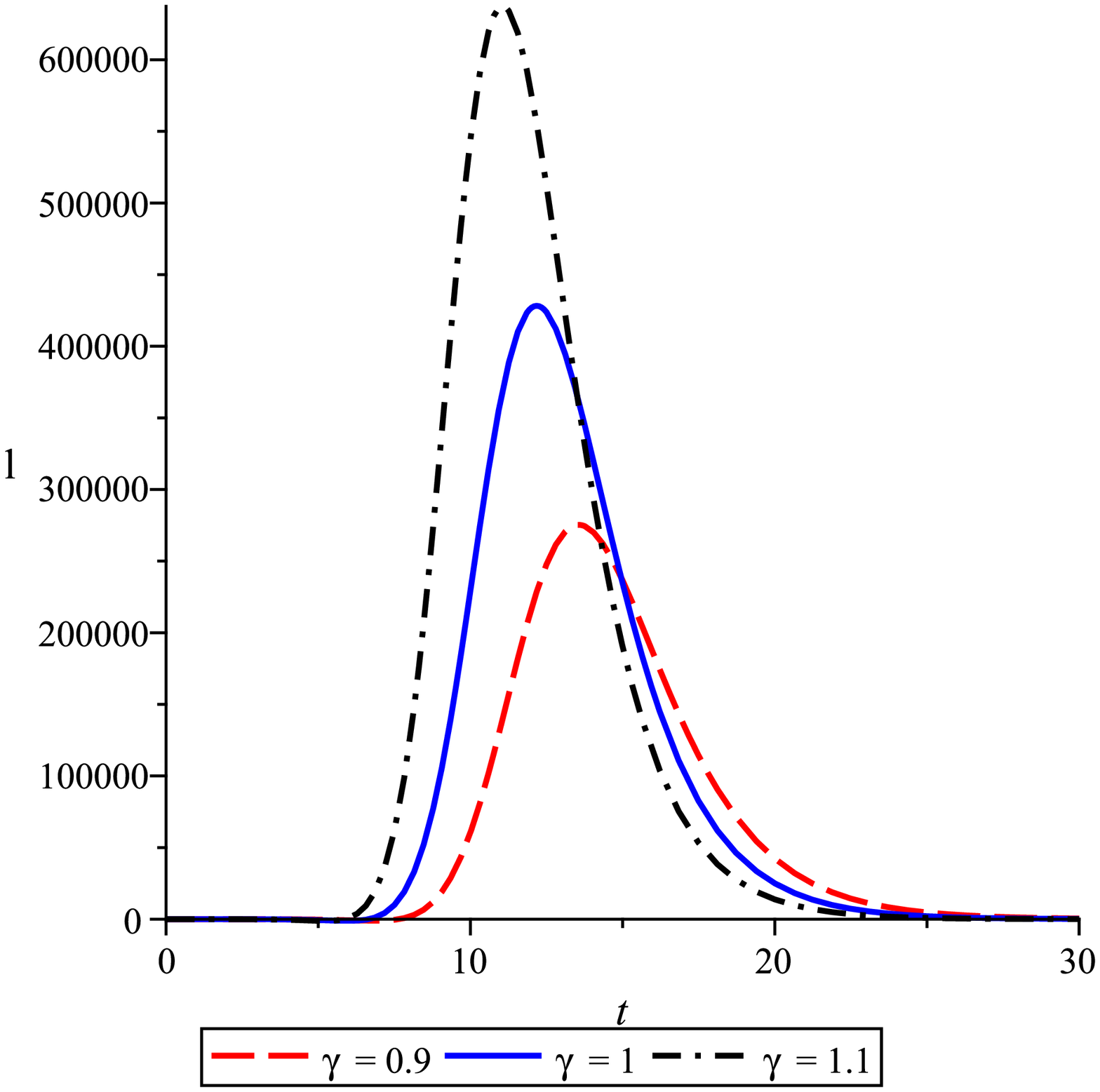}
\caption{The figures show the variation of\\ the ``lerk'' ($l$) parameter throughout the entire\\ evolution of the universe.}
\end{minipage}
\begin{minipage}{0.45\textwidth}
\includegraphics[width= 0.9\linewidth]{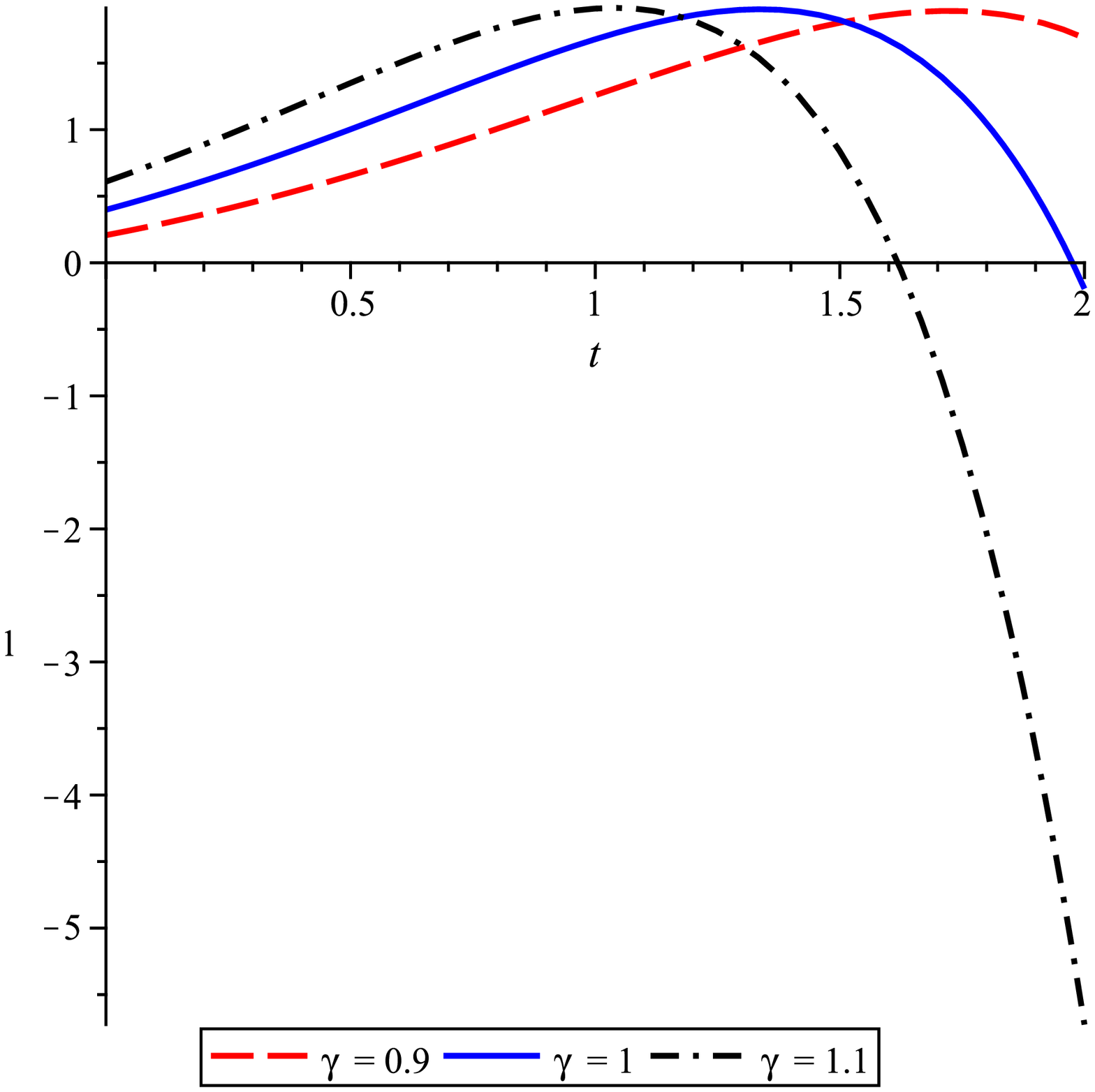}
\caption{The figures are just to show that\\ during the early phase of the universe, `$l$'\\ was $+$ ve.}
\end{minipage}
\end{figure}

In fact, FIG. 8 shows the complete evolution of the jerk parameter `$j$', while
FIG. 9 shows explicitly the variation of $j$ in the early phase of the universe.
There is a transition of $j$ from $-$ ve values to $+$ ve values. The variation of
the snap parameter `$s$' over the entire evolution of the universe has been shown in
FIG. 10, while FIGs. 11 and 12 represent the detailed variation of `$s$' in the early and
late phases respectively. There are three transitions of `$s$' during the whole evolution.
FIG. 13 shows the variation of the lerk parameter `$l$' over the entire cosmic time, and its
variation at early phase is presented in FIG. 14. The figures show two transitions of the lerk
parameter. Finally, the `$m$' parameter is graphically represented
in FIGs. 15--17, which also has the two transitions over the entire evolution.


\begin{figure}
\begin{minipage}{0.45\textwidth}
\includegraphics[width= 1.0\linewidth]{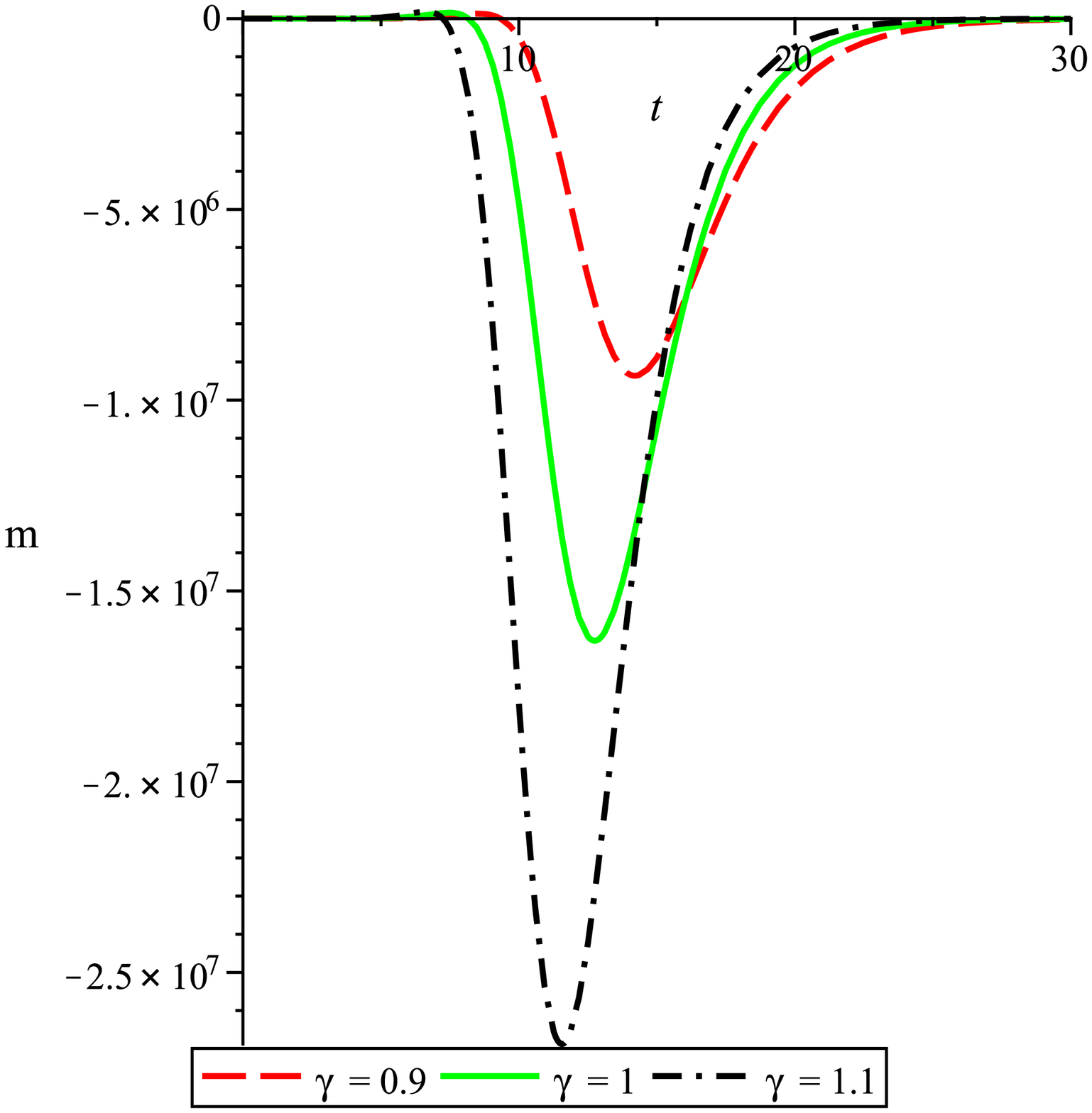}
\caption{The figures show the variation of the\\ `$m$' parameter throughout the entire evolution\\ of the universe.}
\end{minipage}
\begin{minipage}{0.45\textwidth}
\includegraphics[width= 0.9\linewidth]{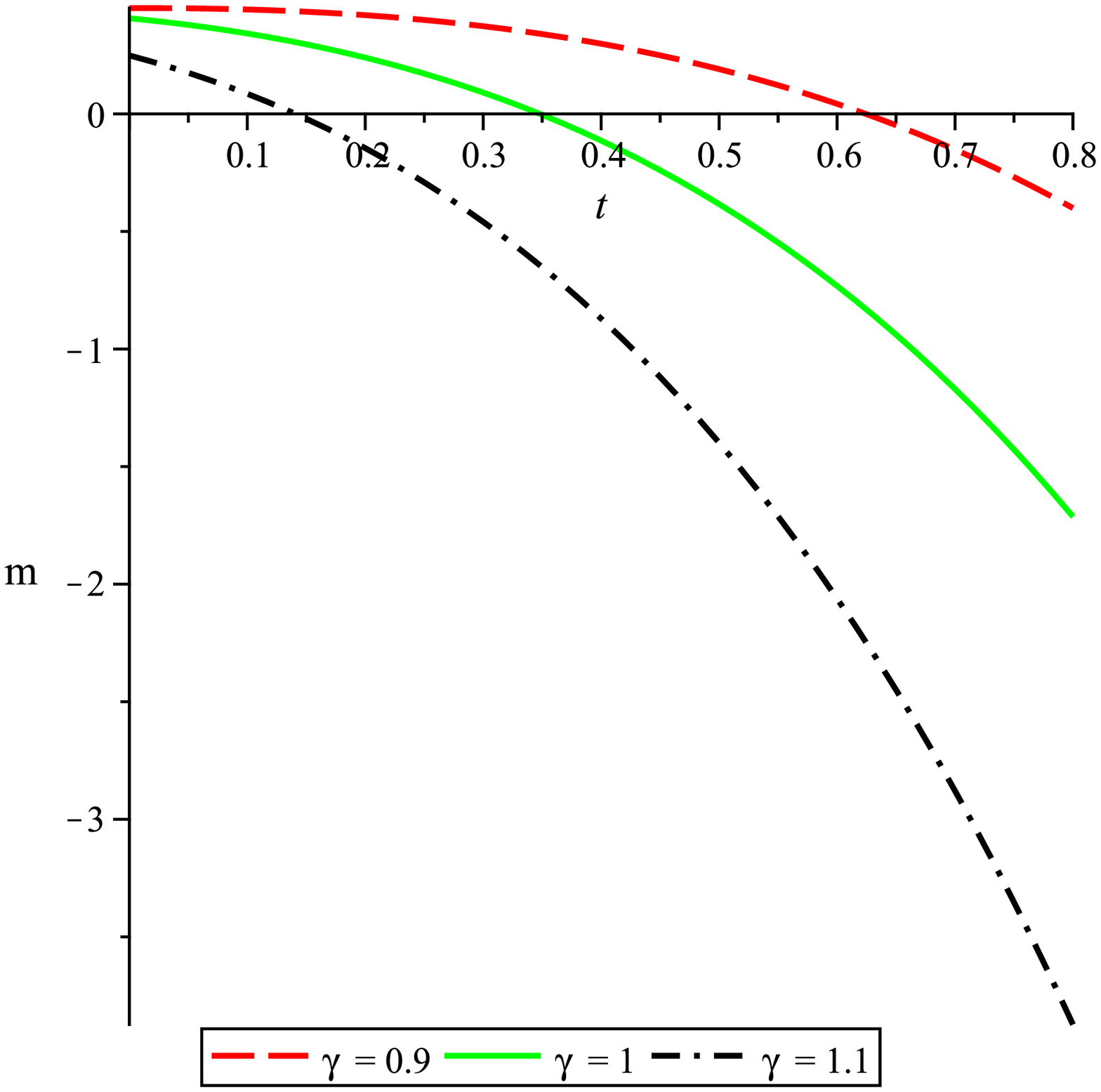}
\caption{The figures are just to show that\\ during the early phase of the universe, `$m$'\\ started from $+$ ve.}
\end{minipage}
\begin{minipage}{0.45\textwidth}
\includegraphics[width= 0.9\linewidth]{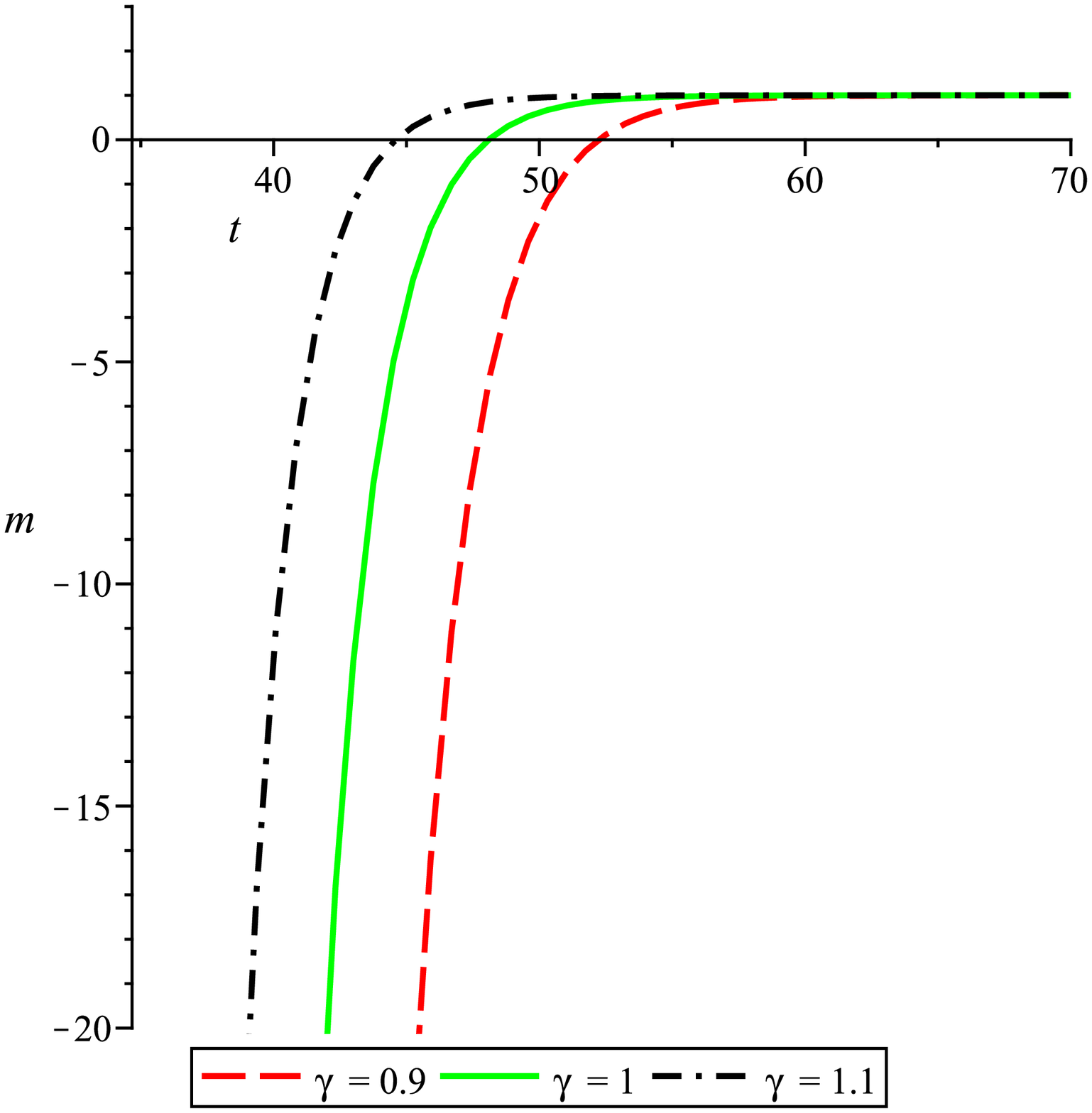}
\caption{This is just to show that when $t \longrightarrow \infty$,\\ `$m$' becomes $+$ ve.}
\end{minipage}
\end{figure}

\section{Field Theoretic Description of Cosmic history}
\label{Field}
In this section, we shall describe the cosmic evolution from the field theoretic
point of view by describing the whole dynamical process as the evolution of a scalar
field $\phi$ having self interacting potential $V (\phi)$, or, equivalently,
the evolution of the present effective imperfect fluid can be described by a minimally
coupled scalar field. Thus, the energy density and the thermodynamic pressure of the cosmic
fluid are given by

\begin{eqnarray}
\rho&=& \frac{1}{2} \dot{\phi}^2+ V (\phi),~~\mbox{and},~~p_{eff}= p+ \Pi= \frac{1}{2} \dot{\phi}^2- V (\phi).\label{sp-scalarfield1}
\end{eqnarray}

Thus, for the present isentropic thermodynamical system, we have

\begin{eqnarray}
\dot{\phi}^2&=& \frac{\gamma}{\kappa} (\mu^2 H- \alpha^2),\label{sp-scalarfield2}\\
V (\phi)&=& \frac{1}{2 \kappa} [6 H^2- \gamma(\mu^2 H- \alpha^2)],\label{sp-scalarfield3}
\end{eqnarray}

where $\Pi$ is eliminated by the isentropic condition equation (\ref{ic}), particle creation rate $\Gamma$ is obtained
from Eq. (\ref{gamma}), and the first Friedmann equation in (\ref{efe}) has been used. Thus, integrating the above equation
(\ref{sp-scalarfield2}), the explicit form of $\phi$ is

\begin{eqnarray}
\phi&=& \phi_0- \frac{4}{\mu^2 \sqrt{\kappa \gamma}} \exp\left(-\frac{\mu^2 \gamma}{4} (t-t_0)\right),\label{sp-scalarfield4}
\end{eqnarray}

and the potential can be expressed explicitly in $\phi$ as

\begin{eqnarray}
V (\phi)&=& V_0+ V_1 (\phi- \phi_0)^2+ V_2 (\phi- \phi_0)^4,\label{sp-scalarfield5}
\end{eqnarray}

where $\phi_0$ is the constant of integration, and, $V_0$, $V_1$, $V_2$ are the constants depending on
$\mu$, $\alpha$, and $\gamma$. Note that, the scalar field has always a value less than $\phi_0$. Further, the particle creation
rate $\Gamma$ can be expressed as a function of $\phi$ as

\begin{eqnarray}
\Gamma&=& \Gamma_0+ \Gamma_1 (\phi- \phi_0)^2+ \frac{\mu^2}{1+ \Gamma_2 (\phi- \phi_0)^2},\label{sp-scalarfield6}
\end{eqnarray}

with $\Gamma_0$, $\Gamma_1$, and $\Gamma_2$ as constants.

It is worthwhile to mention that, in the asymptotic limit (i.e., $\Lambda$CDM era), $\phi$ becomes a
constant ($= \phi_0$), and the potential behaves as the cosmological constant $\Lambda$.

Furthermore, in view of the slow roll approximation during the inflationary phase, the density
fluctuations are of the form $\delta_H \sim H^2/ \dot{\phi}^2 \sim 10^{-5}$ \cite{Peacock1, Basilakos1}. So, for the
present model,  we see

\begin{eqnarray}
\delta_H &\sim& \frac{\kappa H}{\gamma \mu^2 \left(1-\frac{H_0}{H}\right)}\mid_{Initial~epoch:~H= H_I}~~,~~\mbox{i.e.,}\label{sp-scalarfield8}\\
H_I/H_0&\sim&~~\delta_H (\frac{\gamma \mu^4}{\kappa \alpha^2})-1~~\simeq \delta_H (\frac{\gamma \mu^4}{\kappa \alpha^2}).\label{sp-scalarfield9}
\end{eqnarray}

As $\kappa= 8 \pi G= 8 \pi l_{pl}^2$ (in units $\hbar= c= 1$), so, the ratio of the Hubble parameter
at the two accelerating phases (initial and present) is proportional to $l_{pl}^{-2}$. At the
Planck size of the universe, i.e., $l_{pl} = 10^{-35}$ m, the Hubble parameter is $H_I \sim 10^{45}$ sec$^{-1}$ \cite{Zhu1},
so, $H_0 \sim 10^{-23}$ sec$^{-1}$. Thus, the cosmological parameter $\Lambda$ is $\sim 10^{-47}$, the
present observed value.

\section{Hawking like Radiation in the context of particle creation mechanism}
\label{Hawking}
We have made attempts in this section to interpret the mechanism of particle creation
as the phenomenon of Hawking like radiation from the homogeneous and isotropic FLRW space-time
model. Normally, in the particle creation mechanism, the dissipative term behaves as an
effective bulk viscous pressure, and, as a result, there is a negative pressure
term in the Einstein's field equations. Also, it is found that, by proper choice
of the particle creation rate, there is an accelerated expansion of the universe
both at the early stage (inflation), and at late-time, i.e., a complete description
of the cosmic history. So, it is very natural to enquire whether there is any
similarity between this mechanism with Hawking radiation as the inflationary stage
can be described by the Hawking radiation in the FLRW universe \cite{Modak1, Modak2}.

In Hawking radiation, initially due to the enormous size of the black hole, the
evaporation process was very slow, and the process gradually became faster and
faster with the diminision of the size of the black hole. As a result, the
temperature of the black hole also increases with the process. At the end,
when the black hole becomes of Planck dimension, quantum gravity should
come into picture. On the other hand, the evolution of the universe is in the
reverse direction of the black hole evaporation. At the beginning, the quantum
gravity effects are important due to Planck size of the universe. But, with the evolution
of the universe, Hawking radiation comes into picture, and the temperature
gradually decreases.

Moreover, there is another basic difference between these two evolution processes.
In black hole evaporation, the created particles escape outside the event horizon, and,
move to asymptotic infinity, while for evolution of the FLRW, the situation
is just reversed --- the particles created near the (apparent) horizon will
move inside. As a result, due to black hole evaporation, there is a loss of
energy, but the universe gains energy due to particle creations. Furthermore,
due to isotropic nature of the FLRW model, the radiation should be uniform
in all direction, and, we have from the Stefan--Boltzmann radiation law (SBRL)
\cite{ERBJ}

\begin{eqnarray}
P&=& \frac{dQ}{dt}= \sigma A_H T^4,\label{sp-newHR1}
\end{eqnarray}

where $\sigma= \pi^2 \kappa_B ^2/60 \hbar^3 c^2$ is the Stefan-Boltzmann constant,
$T$ is the radiation temperature, $Q$ is the heat radiated by the black body,
$P$ is the net radiated power, and $A_H$ is the
radiating area.\\

Using the above SBRL in the first law of thermodynamics, i.e.,

\begin{eqnarray}
P&=&\frac{dQ}{dt}= \frac{d}{dt} (\rho V)+ p \frac{dV}{dt},\label{sp-newHR2}
\end{eqnarray}

we obtain

\begin{eqnarray}
\dot{\rho}+ 3 H \{(\rho+ p)- \sigma T^4\}&=& 0.\label{sp-newHR3}
\end{eqnarray}

So, comparing with matter conservation equation (\ref{ece}), and considering the thermal
process to be isentropic (i.e., using Eq. (\ref{ic})), the particle creation rate is related to the
temperature as

\begin{eqnarray}
\Gamma&=& \frac{\sigma}{\gamma} \frac{T^4}{H},\label{sp-newHR4}
\end{eqnarray}

Now, choosing $\Gamma$ for the present model (in Eq. (\ref{gamma})), we get

\begin{eqnarray}
\sigma T^4 &=& \gamma(-\mu^2 H+ 3 H^2 + \alpha^2).\label{sp-newHR5}
\end{eqnarray}

In the early phases of the evolution of the universe, $H$ is very large, so, the above equation can
approximately be written as

\begin{eqnarray}
\sigma T^4&\simeq& 3 \gamma H^2= \kappa \gamma \rho,\label{sp-newHR6}
\end{eqnarray}

which is the usual black body radiation. On the other hand,
at late-times, when the universe is very nearly to the
equilibrium configuration, the Clausius relation becomes

\begin{eqnarray}
T\dot{S}&=& \dot{Q}= \mbox{constant}.\label{sp-HR1}
\end{eqnarray}

But, at late-time, the entropy is proportional to `$a$' (see Eqns. (\ref{LambdaCDM1})
and (\ref{LambdaCDM2})), and, hence, $\dot{S} \propto H_0~a$. Thus, from the above relation, we get

\begin{eqnarray}
T&\propto& 1/a,\label{sp-HR2}
\end{eqnarray}

which is nothing but the present CMB temperature. Lastly, it should be noted that,
in the early phase of the universe, when $\Gamma \simeq H$, $T\propto H^{1/2}$, so, the
temperature is not exactly Hawking temperature ($T \propto H$), rather, Hawking-type radiation.
The same is also true for the late-time evolution (where CMB temperature is dominant over Hawking
temperature). Finally, the model becomes very close to the $\Lambda$CDM era; $H$ is found to be a constant,
and, hence, the particle creation rate, as well as, the temperature are constant, and there is no
analogy with Hawking type radiation.

\section{Summary and discussions}
\label{Summary}
Relativistic cosmology with usual perfect fluid (having barotropic equation of state)
as cosmic substratum is considered in the present work in the context of non-equilibrium thermodynamics.
In the framework of particle creation mechanism, dissipative phenomenon is reflected as an effective bulk
viscous pressure. In the second order formulation of non-equilibrium thermodynamics due to Israel and Stewart,
the dissipative pressure acts as a dynamical variable whose evolution is characterized by non-linear
inhomogeneous evolution, however, the entropy flow vector satisfies the second law of thermodynamics. Due to
our inability in solving the non-linear evolution equation, the thermal process is assumed to be adiabatic (i.e.,
the entropy per particle is constant), and, as a result, the dissipative pressure is linearly related with
the particle creation rate. In earlier works \cite{Lima1, Zimdahl2, Abramo1, Gunzig1998, Lima2, Basilakos1, Saha1, Pan1},
the particle creation rate at different cosmic stages had been chosen phenomenologically
(with some basis from thermodynamics), and the solutions
for different physical and thermodynamical variables show a continuity across the transition epochs.
Subsequently, a single particle creation rate \cite{Chakraborty2} describes the whole evolution, but no analytic
solution was found. The present work also considers the particle creation rate as a function of the Hubble
parameter in a phenomenological way which not only describes the cosmic story from inflation
to late-time acceleration, but also, the model has an analytic solution. Also, graphically, we have shown the evolution
of the scale factor, Hubble parameter, deceleration parameter, the thermodynamic variables, namely, the temperature,
entropy, and the number density. It is found that, at late-time, the model asymptotically approaches to $\Lambda$CDM,
and it agrees with the latest Planck data set \cite{Planck2015}. A cosmographic analysis has
also been done for our model, and, the cosmographic parameters, namely, the jerk ($j$),
snap ($s$), lerk ($l$), and the $m$ parameter have been graphically shown
against the cosmic time. From the figures, we conclude that the above parameters
respectively have one, three, two, and two transitions through the cosmic evolution.
Further, from the field theoretic view point, it has been shown that,
it is possible to have a minimally coupled scalar field as equivalent to the cosmic fluid for the description of the
evolution of the universe. In the late-time asymptotic limit, the self similar potential behaves as cosmological
constant. Finally, in resemblance with Hawking radiation, it is found that, at early epoch of the evolution, the
temperature corresponds to black body radiation, while for very close to $\Lambda$CDM, the temperature is related
to the CMBR --- in both cases, the temperature is not the Hawking temperature, rather, Hawking type radiation.


\begin{acknowledgments}

The author S.C. acknowledges the UGC-DRS Programme in the Department of Mathematics,
Jadavpur University. SP acknowledges CSIR, Govt. of India for financial support through
SRF scheme (File No. 09/096 (0749)/2012-EMR-I). The author S.S. is thankful to the
UGC-BSR Programme of Jadavpur University for awarding Research Fellowship.

\end{acknowledgments}


\end{document}